\documentstyle[12pt,aaspp4,epsf]{article}

\begin{document}

\def\intldate{\number\day\space\ifcase\month\or
January\or February\or March\or April\or May\or June\or
July\or August\or September\or October\or November\or December\fi
\space\number\year}

\def \deg    {$^{\circ}$}
\def \eg     {{e.g., }}
\def \cf     {{cf.\ }}
\def \ie     {{i.e., }}
\def \hub    {{$H_{\hbox{\eightrm 0}}$}}
\def \hunits {{km s$^{\hbox{\eightrm --1}}$ Mpc$^{\hbox{\eightrm --1}}$}}
\def \sec    {$^{s}$}
\def \arcsecpoint {$.^{''}$}

\def \DoPhot	{D{\sc o}PHOT}
\def \DOPHOT	{D{\sc o}PHOT}
\def \dophot	{D{\sc o}PHOT}
\def \hi    {\ion{H}{1}}
\def \hii    {\ion{H}{2}}
\def\tightenlines{\def\baselinestretch{1}\small\normalsize}
\def\gtorder    {\mathrel{\raise.3ex\hbox{$>$}\mkern-14mu\lower0.6ex\hbox{$\sim$}}}
\def\ltorder    {\mathrel{\raise.3ex\hbox{$<$}\mkern-14mu\lower0.6ex\hbox{$\sim$}}}
\def\sb         {{\rm mag~arcsec$^{-2}$}}
\def\area       {${\rm deg}^2$}
\def\kpc        {\hbox{\rm kpc}}
\def\kms        {\hbox{$\rm km \, s^{-1}$}}
\def\pc         {\hbox{ pc }}
\def\yr         { \, {\rm yr}}
\def\peryr      { \, {\rm yr^{-1} }}
\def\vlos       { v_{\rm los} }
\def\lsim       {\rlap{\lower .5ex \hbox{$\, \sim \, $} }{\raise .4ex \hbox{$\, < \, $} }}
\def\gsim       {\rlap{\lower .5ex \hbox{$\, \sim \, $} }{\raise .4ex \hbox{$\, > \, $} }}
\def\solar      { {\odot} }
\def\lsolar     { {\rm L_{\odot}} }
\def\msolar     { \rm {M_{\odot}} }
\def\rsolar     { {R_{\odot}} }
\def\rnot               { {R_{o}} }
\def\vsolar     { {v_{\odot}} }
\def\vnot       { {v_{o}} }
\def\surfmunit  { \rm {\, \msolar \, pc^{-2}} }
\def\HI         {{H{\sc I}}}
\def\etal       {{et~al.}}
\def\mags       {{ \, \rm mag }}   
\def\percubicpc { { \pc^{-3} } }
\def\abs        { \hbox{ \vrule height .8em depth .4em width .6pt } \,} 
%
%

%
%

\title{Red Clump Morphology as Evidence Against\\ 
a New Intervening Stellar Population as the\\ 
Primary Source of Microlensing Toward the LMC}
\author{Jean-Philippe Beaulieu and Penny D. Sackett}
\affil{Kapteyn Astronomical Institute 9700 AV Groningen, The Netherlands\\
beaulieu@astro.rug.nl, psackett@astro.rug.nl}

\begin{abstract} 

We examine the morphology of the color-magnitude diagram (CMD) for core 
helium-burning (red clump) stars to test the recent suggestion by 
Zaritsky \& Lin (1997) that an extension of the red clump in 
the Large Magellanic Cloud (LMC) toward brighter magnitudes 
is due to an intervening population of stars that is responsible for 
a significant fraction of the observed microlensing toward the LMC. 
Using our own CCD photometry of several fields across the 
LMC, we confirm the presence of this additional red clump feature, 
but conclude that it is caused by stellar evolution rather than a foreground  
population.  We do this by demonstrating that the feature 
(1) is present in all our LMC fields, 
(2) is in precise agreement with the location of the blue loops 
in the isochrones of intermediate age 
red clump stars with the metallicity and age of the LMC, 
(3) has a relative density consistent with stellar evolution and 
LMC star formation history, and
(4) is present in the Hipparcos CMD for the solar neighborhood where 
an intervening population cannot be invoked. 
Assuming there is no systematic shift in the model isochrones, which 
fit the Hipparcos data in detail, 
a distance modulus of $\mu_{LMC} = 18.3$ provides the best fit to 
our dereddened CMD.
\end{abstract}


\vskip 1cm
\centerline{\it Accepted for publication in the Astronomical Journal}
\newpage


\section{Introduction}

The recent discovery of an overdensity of stars in the 
color-magnitude diagram (CMD) of the Large Magellanic Cloud (LMC) having 
nearly the same color as the ``red clump'' of core He-burning stars 
but extending $\sim$0.9~mag brighter 
has been interpreted as an intervening population of stars at 
$33 - 35$~kpc that may represent a dwarf galaxy  
or tidal debris sheared from a small Milky Way satellite 
(Zaritsky \& Lin 1997, hereafter ZL).  
Zaritsky \& Lin label this overdensity 
the VRC (vertical extension of the red clump), and reject other 
possible explanations to conclude that the VRC represents a 
massive foreground population with about 5\% of angular surface 
density of the LMC itself.  If true, this conclusion 
would have profound consequences for the interpretation 
of Galactic microlensing studies (Renault \etal\ 1997, Alcock \etal\
1997a) since such debris could, in principle, 
be responsible for a sizable fraction of the microlensing signal 
toward the LMC (Zhao 1996, 1998) that is generally attributed to 
microlensing by 
compact objects in the smoothly-distributed halo of the Milky Way itself. 
 
This particular stellar foreground population as an explanation for the LMC
microlensing optical depth has been challenged on several 
grounds. 
The MACHO team find no evidence for a foreground population at $34$~kpc 
in their extensive photometric database, confirming 
the LMC membership of their Cepheids 
(Alcock \etal\ 1997b, Minniti \etal\ 1997). 
They do find an overdensity of stars in a composite 
MACHO $R$ versus $V-R$ color-magnitude diagram (CMD), but conclude 
that the {\it redder\/} color of this feature is incompatible with 
the hypothesis of a foreground clump population.  
(The feature found by MACHO is unlikely to 
be the VRC, but rather another stage of stellar evolution associated 
with the asymptotic giant branch.)
Gould (1997) argues on 
the basis of surface photometry of LMC performed by deVaucouleurs (1957) 
that one of the following is true about any luminous foreground population: 
(1) it does not extend more than 5\deg\ 
from the LMC center, (2) is smooth on 15\deg\ scales, 
(3) has a stellar mass-to-light ratio 10 times that of known populations, or 
(4) provides only a small fraction of the microlensing optical depth.
Using a semi-analytic method to determine the phase space distribution 
of tidal debris, Johnston (1998) has analyzed the Zhao (1998) proposition,  
concluding that an ad hoc tidal streamer to explain the microlensing 
optical depth toward the LMC would cause unobserved overdensities 
of 10-100\% in star counts elsewhere in the Magellanic Plane or would 
require disruption precisely aligned with the LMC within the last $10^8$ 
years.  Bennett (1997) argues that a recently-determined stellar mass 
function combined with the assumption that the putative foreground 
population has a star formation history similar to the LMC results  
in an implied microlensing optical depth from the VRC 
that is only a small fraction of that determined by microlensing 
observations.

We will argue that the VRC feature observed by ZL in color-magnitude 
diagrams of the LMC originates in the LMC itself.  
Using BVR CCD photometry of several fields at different 
locations in the LMC, we confirm the presence of substructure in 
LMC red clump morphology corresponding to the VRC.  
In contrast to ZL, however,  
we argue that the origin is likely to be due to stellar evolution, 
not an intervening population.  
We begin by illustrating that the VRC is seen in all our fields.  
Because the 
red clump morphology varies slightly in color and magnitude over 
the face of the LMC, interpretation of composite CMDs is complicated 
by the superposition of different features.   
We therefore focus on individual LMC fields, 
overlaying isochrones and evolutionary tracks of the appropriate 
metallicity and age in order to demonstrate 
that the VRC corresponds precisely in magnitude 
and color to the so called ``blue loops'' experienced by aging 
intermediate-mass core He-burning stars.   
We then show that similar red clump morphology 
is present in the CMD of Hipparcos, which 
probes stellar populations on scales of $~100$~pc from the Sun, 
where intervening dwarf galaxies or tidal debris cannot be invoked.  
Finally, we analyze the argument used by ZL to reject 
stellar evolution as the cause of the VRC, and show that a more 
realistic model for the star formation history in the LMC is 
not only consistent with the VRC, but also provides a better fit to the 
data.

\section{New BVR LMC Photometry}

In January 1994, Bessel BVR photometry was performed with the Danish 1.5m
telescope at ESO La Silla on the EROS\#1, EROS\#2 and MACHO\#1 microlensing
candidates and a fourth field was taken far from the bar; 
we will refer to these fields as F1, F2, F3 and F4 respectively. 
The detector was a thinned, back-illuminated, AR-coated Tektronix 
$1024 \times 1024$ CCD with a nominal gain of 3.47 e-/ADU,
readout noise of 5.25 e- rms, and pixel size of $24 \mu m$
corresponding to 0.38\arcsec\ on the sky. 
The detector is linear to better than 
1\% over the whole dynamic range and is not affected by any large 
cosmetic defects.  
Observational and field characteristics are listed in Table I. 
The CMD of these fields have been 
used to calibrate data obtained by the EROS microlensing survey, 
further details can be found in Beaulieu \etal\ (1995).  
 
We have performed a reanalysis of these BVR data with \DoPhot\ 
(Schechter, Mateo \& Saha 1993).  Typical \DoPhot-reported errors 
on relative photometry are 0.02 mag at V = 19 
(typical for the clump stars) for the cosmetically superior 
(\DoPhot\ type 1) stars used throughout this analysis.  
Absolute calibration was performed using Graham (1982) 
and Vigneau \& Azzopardi (1982).
Foreground extinction was estimated using 
\hi\ and IRAS maps (Schwering \& Israel 1991);   
these corrections are listed in Table~1 for each field.  
Beginning with this foreground extinction and assuming a metallicity 
of $Z = 0.008$ for the LMC and a helium abundance of $Y = 0.25$, 
we then vary the internal extinction to achieve the best fit 
of the main sequence to the isochrones of Bertelli \etal\ (1994).  
In fields F1 and F4, this produces a total extinction equal to 
that of the foreground (no internal reddening); in fields F2 and 
F3, the internal extinction results in an additional $E(B-V)_{\rm int} = 0.04$. 
We will show that although correcting for extinction is important, 
the difference between the foreground reddening and the reddening 
determined from main sequence fitting does not affect our conclusions. 

Unless otherwise stated, we assume a distance modulus of $\mu_{LMC}=18.3$ 
for the LMC.  
This choice of distance modulus produces the best fits to the 
isochrones along the main sequence and in the red clump.

\subsection{Features in the LMC color-magnitude diagrams}

A calibrated composite $(V-R)-V$ CMD for all four fields is shown in 
Fig.~1 both with and without extinction corrections for the 
individual fields.  
Comparable CMD have been presented and analyzed by 
Vallenari \etal\ (1996, and references therein). 
The red clump is the most notable feature in the LMC 
other than the main sequence itself and 
is clearly visible in the composite CMD at about 
$V_0 \approx 18.7$ and $(V-R)_0 \approx 0.5$ ($(B-V)_0 \approx 0.9$).  
First identified by Cannon (1970), red clump stars 
are the counterpart of the older horizontal branch (HB) in 
globulars, and represent a post helium-flash stage of stellar 
evolution (for a review, see Chiosi, Bertelli \& Bressan 1992).  
Differential reddening may be responsible for the elongated, tilted red clump 
in the dustier bar fields F1 and F3, although   
differences in the age distribution and contamination by the 
red giant branch bump (see \S2.2.4) are also likely to play a role.  
The morphology of the red clump itself may also be shaped by mass-loss 
(\cite{ren88}, \cite{jim97}). 
Our analysis is aimed at understanding that portion of 
red clump morphology relevant to testing the hypothesis 
that the VRC is due to an intervening stellar population. 

\subsubsection{The red clump vertical extension (VRC) and supra-clump}

A narrow vertical extension of the clump 
having nearly the same color as the peak red clump density 
but extending up to $\sim$0.8~mag brighter can also be seen.  
A second source of substructure, or ``supra-clump,'' is apparent 
at a position $\sim$0.8~mag brighter in $V$  
and $\sim$0.1~mag redder in $(V-R)$ than the peak of the primary red clump. 
The positions of these features are marked in Fig.~1.  
This substructure can be seen not only in the composite CMD, but 
also in the calibrated, extinction-corrected $(V-R)-V$ 
CMDs for each of the individual four LMC fields presented in Fig.~2.  
Fields F1 and F3 are located close to the LMC bar; 
fields F2 and F4 further away at about $2 - 3$\deg\ from 
the center of the LMC. The magnitude and direction of the reddening 
vector is shown for each field.  

Examination of Fig.~2 indicates that although the overall morphology  
of the CMD is the same in each field, field-to-field variations 
can be seen in the extent of the red clump and in relative stellar 
densities along the main sequence and between the main sequence 
and the red clump.  
Some of this variation is due to differences in crowding; 
less severe crowding clearly results in deeper photometry 
for the F2 field, for example, and thus a larger number of main 
sequence stars. 
Since histograms reveal a small relative excess of bright main sequence stars 
in the outer compared to the inner LMC fields, the star 
formation history of our fields may also be somewhat different, 
complicating any analysis that rests on a composite CMD drawn from 
several regions of the LMC. 
We therefore choose to analyze the fields independently.

\subsubsection{Comparison of with the VRC of Zaritsky and Lin}

Using contour plots of our four CMD, we determine 
the position of the red clump at peak density  
and the extent of other substructures relative to the clump. 
The results are summarized in Table~2.  
The {\it relative\/} position of the vertical extension is remarkably 
constant in each of our fields and is also consistent with that found by ZL: 
we find that the vertical extension has a $V-R$ color that varies 
no more than 0.03~mag from that of the primary clump peak and that it 
extends at least 0.85~mag brighter in $V$ than the red clump peak. 
The second, redder substructure also maintains a constant 
relative position to the red clump from field to field. 
To within 0.02~mag in every field, this supra-clump is 0.85~mag brighter in $V$ 
and 0.10~mag redder in $V - R$ than the peak density of the red clump.

To test whether the relative stellar density in our CMD 
within the region of the VRC is consistent with that found by ZL, 
we return to our composite and individual dereddened $(V-R) - V$ CMD.
In color-magnitude space we define the 
vertical extension ($0.4 \leq (V-R) \leq 0.5$, $17.7 \leq V \leq 18.3$), 
the supra-clump ($0.5 \leq (V-R) \leq 0.65$, $17.7 \leq V \leq 18.3$) 
and primary clump ($0.4 \leq (V-R) \leq 0.6$, $18.4 \leq V \leq 19.5$) 
and count the stars within these regions.  
These counts are summarized in Table~3, indicating that the VRC represents 
$\sim$8\% and the supra-clump $\sim$14\% of the primary clump in the
composite CMD. In individual CMDs, the VRC to red clump fraction VRC/RC 
varies from 6\% (in fields F1 and F3) to 15\% in field F2. 
The supra-clump to red clump fraction SC/RC varies from 12\% to 16\%.
Slightly different choices for the relevant boxes, for example 
narrowing them to reduce contamination from stars in other stages 
of stellar evolution, yield very similar fractions.  
Since our estimate from the composite CMD for the VRC/RC fraction 
has an uncertainty of $\sim \pm 1\%$ from counting statistics alone, 
it is consistent with the ZL estimate of 
$\leq 7\%$ for the relative projected angular surface density of the VRC. 

To summarize, 
the characteristics that we measure for vertical extension to the red 
clump and the redder supra-clump peak are identical 
within the uncertainties to those found by ZL and the MACHO team 
respectively.  We therefore identify the vertical feature seen in our data 
with the VRC discussed by ZL and the redder supra-clump with 
the overdensity discussed by Alcock \etal\ 
In the following section we propose stellar evolutionary origins for 
each of these features.  

\subsection{Analysis of the Individual CMDs}

The position of red clump stars in a color-magnitude diagram depends on 
their mass, and thus their age, as they pass through 
this evolutionary stage.  Using the isochrones of Bertelli \etal\ (1994), 
we plot in Fig.~3 the mean locus of the core Helium-burning phase 
for a variety of stellar ages in a $(B-V) - V$ CMD.  
This time-averaged locus marks where stars in core Helium-burning phase 
are likely to be found.  
We have used tracks with metallicity and helium abundance 
appropriate to the LMC, namely $Z=0.008$ and $Y=0.25$. 
Note that clump stars with ages in the range $\log({\rm Age})=8.6-9$ 
(0.4 to 1 Gyr) exhibit color changes smaller than $0.03$mag in $(B - V)$
while differing in $V$~magnitude by 0.79~mag.  We now compare these  
theoretical expectations with the VRC in our own LMC data.

\subsubsection{Choice of Distance Modulus}

We begin by focusing on the field F2, since the low level of 
crowding in this outer field has resulted in the best photometry of 
our four fields.    
The luminosity function for main sequence stars 
in this field peaks at about $V = 21.5$; 
in the bar fields this occurs about 0.5~mag sooner.  
The $(B-V) - V$ CMD for field F2, dereddened by $E(B-V)=0.12$ 
is shown in the upper panel of Fig.~4, for two different choices 
for the LMC distance modulus.  
Overplotted are theoretical isochrones from Bertelli \etal\ (1994)  
computed with new radiative opacities (OPAL) for 
metallicities appropriate to the LMC ($Z = 0.008$ and $Y = 0.25$) and 
ages corresponding to 0.25, 0.40, 0.63, 1.0 and 2.5 Gyr 
(log(Age) = 8.4, 8.6, 8.8, 9.0 and 9.4). 
The lower panel enlarges the region of the CMD near the red clump region, 
which is now plotted as contours under the mean locus of the core
He-burning phase from Fig.~3.   Both the fit to the main sequence 
and the red clump is significantly improved for the smaller distance  
modulus of $\mu_{LMC} = 18.3$.  Since this improvement was apparent in 
all our fields, we fixed $\mu_{LMC}$ at this value.    

\subsubsection{Importance of Dereddening}

Using the same isochrones from Bertelli \etal\ (1994),  
ZL concluded that stars of age 2.5 Gyr provided 
the best fit to the red clump morphology seen in 
their LMC data.  If, following ZL, we do not apply an  
extinction correction, the 2.5~Gyr isochrone 
does indeed provide a plausible fit to the red giant branch.  
With our extinction correction, however, 
this isochrone falls at the very reddest (or oldest) edge of 
the red clump in field F2.  
As Fig.~5 illustrates, this is true for all of our fields.  
Note that if we had applied only foreground extinction corrections 
due to the Milky Way itself, the discrepancy with the 2.5~Gyr 
isochrone would still be present: for two fields the internal 
extinction (as determined by our main sequence fitting) is negligible, 
for two others $E(B-V)$ is increased by only 25 - 33\%. 
Furthermore, younger 
isochrones fit the red clump of each field similarly despite the 
fact that different external extinction corrections (based on 
estimates from \HI\ and IRAS data) were made. 
We take this as an indication that these extinction corrections 
are reasonable and necessary for the proper interpretation of 
the stellar evolution. 

\subsubsection{Identification of the VRC with young He-core burning stars}

The He-core burning phase is associated with 
the horizontal branch in old, metal-poor globulars, but in systems 
such as the LMC containing stars with a variety of ages and metallicities 
the horizontal branch becomes blurred into the red clump in CMD. 
Furthermore, as Fig.~3 makes clear, the horizontal 
branch actually becomes {\it vertical\/} for core He-burning 
stars of intermediate masses and ages between $\sim$0.4 and $\sim$1.0 Gyr. 
Stars of these masses experience the well-known ``blue loops'' 
(see \eg\ Sweigart 1987, Chiosi \etal\ 1992) that are caused by the 
increasing temperature of the outward expanding H-burning shell 
as the He-burning core gains mass.   When the hydrogen in the 
shell is exhausted and helium begins to burn in the shell, the star 
moves redward again in the CMD to quickly join the asymptotic giant 
branch (AGB).  While burning helium in their cores and hydrogen in their 
envelopes, stars spend most of their time near the bluest end of the 
blue loops.  The position of the most 
blueward extension of the loops for $\sim 2-3 \msolar$ stars 
differs substantially in luminosity, but very little in color 
(Fagotto \etal\ 1994); in CMD stars of these masses thus create a vertical 
extension brightward of the blue end of the red clump.   

As can be seen in Fig.~5, all of our fields contain main sequence 
stars as young as 250~Myr.  Since their lifetime on the main sequence 
is relatively short, intermediate mass stars should also be passing 
through the core He-burning phase.  
Indeed, stars with ages between $\sim$0.4 -- 0.8$ \, $Gyr 
lie on a sequence of blue loops whose densely populated 
blue edge corresponds to the position of the VRC.   This is demonstrated  
clearly in Fig.~6, where the time-weighted centroid of core 
He-burning stars from Fig.~3  is overplotted on contours of the 
stellar density for each of our fields. 
The locus of core He-burning stars agrees with the position 
of the primary red clump and the vertical extension.  
The agreement is remarkable, especially considering that we 
have not adjusted the metallicity of the isochrones to achieve a 
better fit.  
We therefore identify the VRC with intermediate mass stars in the LMC 
currently undergoing core He-burning.  In section \S2.3 we show 
that the relative density of stars in the VRC is also 
consistent with a stellar evolutionary origin.

\subsubsection{Identification of the supra-clump with the asymptotic giant branch}

The spatial coincidence of the red and asymptotic giant branches with the 
supra-clump seen both in our data and in the composite CMD 
of the MACHO database (Alcock \etal\ 1997b) suggests that 
this second substructure in the CMD may be associated with giant evolution. 
Indeed, Zaritsky \& Lin (1997) identify the supra-clump 
with the so-called ``red giant branch bump'' 
(RGBB, see \eg Rood 1972, Sweigart, Greggio \& Renzini 1989, 
and Fusi Pecci \etal\ 1990).  
During the red giant branch (RGB) phase, 
color and luminosity evolution pauses as the H-burning shell passes 
through a discontinuity left by the maximum penetration of the 
convective envelope. This pause in the ascension 
results in an overdensity along the RGB. 
Using an LMC distance modulus of 18.3, the apparent magnitude of 
RGBB stars of 2$\msolar$ and LMC metallicities 
(Sweigart, Greggio \& Renzini 1989) agrees with the 
position and extent of the supra-clump in our data.  
The same is true using the parameterized models of Fusi Pecci 
(1990).  The position of the RGBB is however quite sensitive to 
stellar age; older stars be fainter when in the RGBB phase.  
The mean age (and metallicity) of the stars will thus 
determine the position of the RGBB for a mixed stellar population.  

Recent simulations based on isochrones 
from OPAL opacities indicate that the RGBB may lie at the same 
magnitude as the red clump itself in the LMC (Gallart \& Bertelli 1998) 
and thus significantly below the position of the observed supra-clump.   
An alternative explanation for the supra-clump is offered by 
these simulations: a stalling of the evolution at the base of the asymptotic 
giant branch (AGB) as the star struggles to reach thermal equilibrium 
after the helium exhaustion in the core (Gallart 1998). 
This early phase of AGB evolution creates an over-density or bump 
(denoted the AGBB by Gallart) in the CMD since subsequent evolution 
proceeds much faster along the AGB.

Both secondary features in the CMD are thus explained by 
stellar evolution: the VRC by young stars of intermediate-mass 
experiencing core He-burning and the supra-clump by a relatively 
long-lived phase of He-core exhausted stars at the 
base of the asymptotic giant branch.

\subsection{Relative Stellar Density of the VRC}

The position of the vertical extension to the red clump corresponds 
to the end of the blue loops experienced by intermediate age stars 
in their He-core burning phase.  This argues against the need for 
an intervening population to explain the VRC, but is the relative 
stellar density of the VRC consistent with that expected for 
a stellar population with the star formation history of the LMC?

Comparisons of the stellar density in different 
parts of the CMD must be done with extreme care due to the 
selection effects induced by crowding.  In addition, comparing 
regions of the CMD that vary significantly in age increases the 
chances of error due to uncertainties in the star formation history 
and the increase of metallicity with decreasing age.  For example, 
comparing the stellar density of the VRC to that of the primary clump 
not only suffers confusion from the overlapping red giant branch, 
but also from the tracks of low metallicity old (low mass) clump 
stars that intersect those of higher metallicity younger (higher mass) 
clump stars. For these all 
reasons, we choose to compare the stellar density of stars in the VRC 
with stars on the main sequence of comparable age and brightness.  
Our goal is to determine whether simple, reasonable assumptions about 
the IMF and star formation history of the LMC can reproduce not only 
the position of the VRC, but its relative stellar density as well.

\subsubsection{Choice of IMF and Star Formation History}

Detailed models for the stellar density in a particular region of the 
CMD must combine stellar evolution with assumptions about the star 
formation history and the initial mass function (IMF) of the population.  
The LMC is generally believed to be forming stars continuously 
and to have undergone more star formation in the last few Gyr 
than previously (see Olszewski, Suntzeff \& Mateo 1996 for a review), 
both on the basis of ground-based studies 
(Bertelli \etal\ 1994, Girardi \etal\ 1995, and Vallenari \etal\ 1996) and 
deeper studies using Hubble Space Telescope (HST) CMD 
(Gallagher \etal\ 1996, Holtzman \etal\ 1997, Elson \etal\ 1997, and 
Geha \etal\ 1997). 

As Fig.~7 demonstrates, 
in each of our four fields, a constant star formation rate 
with a Salpeter IMF (Salpeter 1955) described by 
$dN = A~M^{-\alpha}~dM$ with $\alpha = 2.35$ 
provides a good approximation to the 
relative densities of stars on the main sequence down to the 
completeness limit of $\sim$18.5 -- 19.  
Field F4 appears to contain a slight excess of very bright main sequence 
stars (Fig.~5 \& 7), which may be indicative of a recent burst of star 
formation activity. 
These results are consistent with ground-based and deeper HST studies 
of the LMC, although most recent work indicates that a slightly 
steeper IMF provides a better match to the luminosity function at 
the expense of some relative density ratios 
(\cite{val96}, \cite{hol97}, \cite{geha97}). 
For the order of magnitude 
estimates presented here, we will therefore assume that star formation has 
proceeded at some relatively constant and arbitrary rate between the epochs 
$8.4 \leq {\rm log(Age)} \leq 9.0$ ($0.25 \leq {\rm Age} \leq 1$ Gyr), 
with $1.85 \leq \alpha \leq 2.85$. 
Because we will count stars on the main sequence and in the VRC corresponding 
to the same initial mass, our density analysis is 
actually independent of the IMF as 
long as it remains unchanged over this same period.  
Only when we examine the effects of crowding will slope $\alpha$ of the 
IMF play a role. 

\subsubsection{VRC and main sequence counts for intermediate mass stars in the LMC}

According to the models of Bertelli \etal\ (1994), 
stars with initial masses between 2.6 and 3.0$\msolar$ will be $\sim$500~Myr 
old in the red clump where they will have the colors and magnitudes 
that place them in the center of the VRC.  Regardless of its slope, 
if the IMF is constant over the last Gyr in the LMC, then the ratio 
of density of these stars in the VRC to their 
density on the main sequence will depend only on the ratio of lifetimes 
in these two phases.  Here, we define the lifetime in the VRC to 
be the length of time that these $2.6 - 3.0 \msolar$ stars burn 
helium in their cores and have $0.4 \leq V-R \leq 0.5$.
We use the evolutionary models of Fagotto \etal\ (1994) to estimate 
the lifetime on the main sequence, which we define as the time during which 
the fraction of hydrogen in the core is above 0.1\%.  
This definition is operationally convenient since it allows us to 
count all stars blueward of the subgiant branch.  With these definitions, 
the ratio of VRC to main sequence lifetimes --- and thus the ratio 
of corresponding VRC to main sequence counts --- is 23\%. 

Using these definitions, stars of initial mass in the range $2.6 - 3.0 \msolar$ 
have $ -0.15 < V-R < 0.15$ and $19.15 < V < 19.5$ on the main sequence, 
and $17.8 \leq V \leq 18.5$ in the VRC where their colors are defined 
as above.  Counts in these regions of the CMD yield VRC to 
main sequence fractions of 98\%, 42\%, 61\%, and 46\%  
for fields F1 through F4, respectively.  Note that the bar fields 
appear to have the highest VRC fraction.  
However, since all of our fields 
are complete at the faintest end of the VRC, but none are complete 
at $V = 19.5$ on the main sequence, these ratios are overestimated 
and must be corrected for incompleteness. 
To estimate the completeness of each field, we compare 
the actual number of stars found in the main sequence bin 
to that predicted by a Salpeter IMF normalized to provide good fits 
to the brighter portion of the main sequence histograms.  
(For this purpose, main sequence was defined empirically to 
mean all stars with $0.2 \leq (V-R) \leq (V-15)^2/80$.) 
The estimates of the completeness {\it C$_{MS}$\/} 
at magnitudes where $2.6 - 3.0 \msolar$ main sequence stars would be found 
are given in Fig.~7 for each field; they range 
from 42\% to 61\%.  It is difficult to estimate the uncertainty in 
these completeness estimates, but they are more dominated by counting 
statistics, which are typically on the order of 15\%,  
than by choice of bins.  
The resulting VRC to main sequence fractions,  
corrected for incompleteness, are 59\%, 25\%, 27\% and 19\% respectively; 
all except bar field F1 are in 1.5$\sigma$ agreement with the 
prediction of 23\% based only on the assumption that the  
star formation rate and IMF slope have remained constant over the last 
1 Gyr in the LMC.   
The composite completeness-corrected VRC to main sequence ratio
is 30\%; if the anomalously high field F1 is excluded, this drops 
to 25\%.  

The completeness correction assumed a Salpeter slope of 
$\alpha = 2.35$.  A steeper IMF would place more stars on the 
fainter part of the main sequence and would thus result in larger 
completeness corrections and smaller VRC ratios.  The converse is 
true for shallow IMF slopes.  How well does stellar evolution fare 
at predicting the 23\% VRC to main sequence ratios for $2.6 - 3.0 \msolar$ 
stars if the IMF is modified?  Redoing the completeness analysis 
with a shallow IMF slope of $\alpha = 1.85$ yields a composite VRC to 
main sequence fraction of 38\%.  A steeper slope of $\alpha = 2.85$, 
as preferred by many LMC luminosity function studies (
Vallenari \etal\ 1996, and references therein, Holtzman \etal\ 1997, 
and references therein), yields a 
smaller value of 21\%, in excellent agreement with the predicted 23\%. 
Finally, note that stars with masses of 
$\sim$4~$\msolar$, corresponding to ages of 200~Myr in the core He-burning 
phase evolve so quickly that they would not be expected to be 
detected in the clump region of the CMD.  
This may explain the upper magnitude cutoff of the VRC.   

In conclusion, stellar evolution combined with a Salpeter IMF and 
constant star formation over the last 1~Gyr can account for $\sim$75\% 
of the observed stellar density in the VRC.   
For the somewhat steeper slopes preferred by most recent LMC studies, 
the fraction rises to 100\%.   
Given the simplicity of the assumptions, counting procedure and  
incompleteness corrections, this agreement is remarkable and leaves  
little room for an intervening population of non-LMC stars in this 
region of the LMC color-magnitude diagram.
  
\section{The Red Clump Vertical Extension in the Hipparcos CMD}

The parallaxes obtained by the Hipparcos mission have allowed the 
determination of distances to stars brighter $M_V = 8$ with accuracies 
on the order of 10\% (Perryman \etal\ 1997).  This has resulted in a more 
accurately-determined color-absolute magnitude diagram for stars within 
$\sim$100 pc than has heretofore been possible, 
resulting in the first clear detection of ``clump giants'' in the solar 
neighborhood (Perryman \etal\ 1995).  
Paczy\'nski \& Stanek (1998) conclude from their analysis that the 
Hipparcos red clump has a mean distance of 105~pc and a geometrical 
mean distance of 98~pc, making it unlikely that extinction 
significantly affects the magnitude or color of the clump.

In Fig.~8, we show the color-absolute magnitude diagram for 16229 stars 
from the Hipparcos catalog with relative distances determined to 
within 10\% and colors determined to within 2.5\%
(Perryman 1995, and references therein).  The red clump is 
clearly visible as a highly concentrated collection of stars with 
a density peaking near $V=0.8$ and $B-V=1.0$.  The fainter, redward tail 
of the clump seen most clearly in the contours may be due to 
the evolving core He-burning stars overlapping the 
giant branch (\ie pre-helium flash giants). 

Also visible in the Hipparcos CMD is a vertical extension of the red clump 
toward brighter magnitudes; its color is indistinguishable from 
that of the peak of the red clump.  Centered near $V=0$, this 
feature is clearly discernible by eye 
at $-0.5 < V < 0.5$ and $0.9 < B-V < 1.1$. 
Its presence and location is robust to a variety of different 
choices for the smoothing and contouring of the CMD.  
A second overdensity at similar magnitudes 
($0.6 < V < -0.6$) but redder colors ($1.2 < B-V < 1.4$) 
is also apparent, and may correspond to the position of the AGBB. 

The isochrones overplotted in the upper panel of 
Fig.~8 assume a solar metallicity of 
$Z=0.02$ for stars of ages 0.4, 1.0, 2.5, 6.3 and 10.0 Gyr.  
It thus appears that the solar neighborhood has undergone 
some star formation in the last 400~Myr, but that the bulk of the stars 
in the red clump in the solar neighborhood are older than those of 
the LMC.  The horizontal extent of the Hipparcos red clump is due 
to older stars that begin burning helium in their cores at fainter, redder 
positions in the CMD and experience less severe blue loops.  
In the lower panel of Fig.~8, the area around the clump is enlarged 
and the stellar density plotted as contours.  The VRC in the 
solar neighborhood can be seen clearly. 
Both the isochrones overplotted in the upper panel of Fig.~8 and the 
theoretical locus of core He-burning stars (of appropriate metallicity) 
overplotted in the lower panel make it clear that the VRC seen in 
the Hipparcos data is due to younger clump stars with 
ages between 300~Myr and 1~Gyr.

\section{The CMD Observations used by Zaritsky and Lin}

Both our CMD and those of ZL exhibit the same vertical extension 
to the red clump, yet we reach different conclusions as to 
the origin of the VRC 
despite using the same isochrones from Bertelli \etal\ (1994) 
with the same metallicity $Z = 0.008$ and helium abundance $Y = 0.25$. 
Why do our conclusions differ?  
ZL choose an isochrone with an age of log(Age)=9.4 as the best match 
to the red giant morphology of their un-dereddened CMD.  
They then note that although younger isochrones with log(Age)=8.6 
can reproduce the increased luminosity of 0.9~mag observed in the VRC, 
the evolutionary models of Bertelli \etal\ predict a 
difference in color of the mean locus of the core He-burning phase
of $\Delta(B-V) = 0.13~$mag, $\Delta(B-I) = 0.23$~mag 
between these two isochrones. 
Since they observe a color difference $\leq 0.07$ in $B-I$ 
between the VRC and the centroid of the red clump, they dismiss  
stellar evolution as the origin of the VRC.  They further note 
that the maximum change in luminosity predicted by Sweigart (1987) 
for stars evolving in the clump is 0.6~mag compared with the 0.9~mag 
that they observe in their CMD.

If the CMD are first dereddened before comparing to model isochrones, 
and the LMC is assumed not to be coeval but to have resulted from 
star formation that has been relatively constant over the last 
few Gyr, one reaches a different conclusion.  Fig.~3 shows how the 
mean $V$ magnitude of the core He-burning clump varies as a function 
of its $B - V$ color for stars of different ages using the models 
of Bertelli \etal\ (1994).  
Indeed, as noted by ZL, a measurable color difference is expected 
between stars of log(Age)=8.6 and log(Age)=9.4.  However, as demonstrated 
by our Figs.~4 and 5, stars as old as 2.5 Gyr (log(Age)=9.4) do not 
fit the dereddened red clump at all, and for some of our fields appear to be 
too red to fit the bulk of red giant branch as well.  
Dereddening by even 0.12~mag 
(comparable to the average {\it foreground\/} reddening of our LMC fields) 
has a dramatic effect on the conclusions since this shifts the mean 
age of the clump to younger ages for which the luminosity of the blue loops 
is a very strong function of age, while the color does not change 
appreciably.   Thus, allowing for a small range in the ages of 
young stars in the red clump is sufficient to reproduce the vertical extent 
of the VRC without significantly modifying its color.  
Note that this color insensitivity of the blue loops to age 
for stars between about 400~Myr and 1~Gyr is also true for the color 
defined by ZL ($C \equiv 0.565*(B- I) + 0.825*(U - V + 1.15)$, as 
shown in Fig.~9 for regions of the CMD comparable to that shown 
by ZL in their Fig.~1.

Zaritsky \& Lin note that if the foreground population hypothesis is 
correct for the origin of the VRC, then the entire LMC CMD should 
show traces of a parallel CMD shifted by 0.9~mag.  They can find no 
such traces in their own data, but do point to the HST color-magnitude 
diagram of Holtzman \etal\ (1997) as corroborating evidence that such 
a parallel feature may be present in the lower main sequence of the 
LMC.  They rightly note that binary stars would be a natural explanation for 
this excess of stars displaced by about $-0.8$~mag from the primary 
main sequence, but dismiss such an explanation as unappealing since 
it cannot simultaneously explain the VRC, where fine-tuning 
would be required to ensure that both members of the binary 
arrive in this region of the CMD at the same time.  
ZL conclude that if binaries are invoked as an explanation for 
the excess population in the lower main sequence, 
another explanation is necessary for the VRC.  While we make no 
claims as to the origin of a parallel main sequence in the LMC, 
we do proffer stellar evolution as the required alternate 
explanation for the VRC.

\section{Conclusions}

Using  BVR color magnitude diagrams obtained in four different regions
in the LMC, we confirm the presence 
of a vertical extension of the red clump having the same color 
as the clump peak, but extending to brighter magnitudes, 
as first mentioned by Zaritsky \& Lin (1997). 
Unlike ZL, however, we conclude that this feature is due to 
stellar evolution, not a foreground population.  Our argument is 
based on noting that the feature 
(1) is present in all our LMC fields, 
(2) is in precise agreement with the time-averaged locus 
the blue loops that represent a relatively long-lived 
phase in the stellar evolution of younger core He-burning stars 
in the clump,
(3) has a relative stellar density consistent with the density on 
the main sequence and the assumption of continuous and constant 
star formation history in the LMC over the past 1 Gyr, and 
(4) is present in the solar neighborhood as demonstrated by the 
Hipparcos color-magnitude diagram.   

The difference in our conclusions despite using 
the same model isochrones (Bertelli \etal\ 1994) 
rests on differences in our reddening corrections 
and assumptions about the ages of LMC stars.  ZL perform no 
reddening correction, whereas we correct for reddening using 
well-determined foreground estimates with a small correction for
internal reddening based on main sequence fitting.  
This results in an average inferred $E(B-V) = 0.14$, 
only 0.02~mag greater than the average deduced for foreground 
extinction alone.  
ZL consider a single age of 2.5~Gyr for the clump with one possible burst 
at 400~Myr; we consider a population more distributed in age 
and weighted toward younger ages than ZL.  
An isochrone of 2.5~Gyr does not provide a good 
fit to the red clump in any of our fields, whereas 
the presence of younger stars in all our LMC fields is clearly demonstrated by 
comparison of isochrones with the position of the clump and the 
stellar density of the upper main sequence.  
We find that stars of ages $\sim 0.4 - 0.6$~Gyr are 
responsible for nearly the full extent of the VRC.

The best fit to both the main sequence and red clump of our CMD 
require a distance modulus to the LMC of $\mu_{LMC} = 18.3$, assuming 
that the agreement of evolutionary models to the CMD of 
solar neighborhood can be taken to imply that no large systematic 
offset in luminosity is present in the theory.

With our justified assumption of continuous and constant star 
formation over the last Gyr in the LMC, current understanding of stellar evolution predicts with precision not only the position of the VRC 
seen in both our data and those of ZL, but also its relative stellar 
density.  
The same stellar evolutionary models (with appropriate 
assumptions for solar-metallicities) reproduce the VRC feature 
in the solar neighborhood CMD measured by the Hipparcos satellite.  
These models when applied to the LMC with a Salpeter IMF can 
account for $\sim$75\% of stellar density of the VRC compared to that 
of the upper main sequence.  A slightly steeper IMF slope, such as 
preferred by many recent luminosity function studies of the LMC, 
reproduces the observed VRC stellar density exactly.
Even with the Salpeter IMF, only one of our four fields (F1) contains 
VRC stars 1.5$\sigma$ in excess of expectations from stellar evolution 
and simple assumptions for the star formation history of the LMC.  
Since microlensing events are currently being reported both in 
the bar and outer regions of the LMC, the 4$\sigma$ discrepant bar field 
alone cannot contain a significant fraction of the microlensing 
optical depth to the LMC in the form of intervening stellar population. 
Indeed three of our four fields (F1, F2 and F3) 
contain a known microlensing candidate and thus have 
non-negligible optical depth to microlensing. 

We conclude that no foreground population need be invoked 
to explain the presence of the vertical extension to the red clump 
in the LMC.  If present, such a population is unlikely to account for more 
than $\sim$25\% of the VRC since the presence of intermediate mass LMC 
stars in the VRC are {\it required\/} by stellar evolutionary 
models in the observed numbers.  
Whatever the primary source of the measured microlensing optical depth 
toward the LMC, it is unlikely to be due to a new foreground population 
that has made its presence evident in this vertical extension of the red 
clump.


\acknowledgments 

We thank Konrad Kuijken for useful discussion and a prompt yet 
careful reading of the manuscript.  We are grateful to 
Andrew Cole for pointing out an error in an early version of this work 
and to Carme Gallart and Alvio Renzini for discussions about the AGB.  
We also thank  
Eric Maurice and Louis Pr\'evot for advice on calibration procedures and 
Richard Naber for with \DoPhot\ subtleties.
This work is based on observations carried out at the European Southern 
Observatory, La~Silla.  Work by PDS while at the Institute for 
Advanced Study in Princeton was supported by the NSF grant AST-~92-15485.


\noindent Note Added in Proof:  The red clump is an increasingly 
popular distance indicator; two recent studies also 
find a smaller LMC distance (Udalski \etal\ 1998, Stanek \etal\ 1998).




\newpage


\section*{Figure Captions}

\noindent
Fig.~1 --- Calibrated, composite $(V-R) - V$ 
color-magnitude diagrams for all four LMC fields described in the 
text. Only the 16,139 stars with cosmetically 
superior point spread functions (\DoPhot\ type 1) in $V$ and $R$ are plotted.  
Both uncorrected (left) and dereddened (right) CMD are shown. 
Positions of the red clump (RC), vertical extension to the 
red clump (VRC), and the supra-clump
are indicated.

\noindent
Fig.~2 --- Calibrated, dereddened $(V-R) - V$ 
color-magnitude diagrams for the individual 
four LMC fields.  Only stars with cosmetically 
superior point spread functions in $V$ and $R$ are plotted.  
The number of stars plotted is 
3614, 4241, 5591, 2693 in $V - R$
for fields F1 through F4 respectively.
\bigskip

\noindent
Fig.~3 --- Centroid of the $V$-band luminosity of the core He-burning phase 
as a function of $B - V$ for stars of differing ages with LMC metallicity 
and helium abundance of Z=0.008 and Y = 0.25 (Bertelli \etal\ 1994).  
Note that no significant color change occurs for stars with ages 
between 400~Myr and 600~Myr; these stars correspond to the bulk VRC in the 
LMC.  A similar sequence for Z=0.02 and Y=0.28 (appropriate to 
the solar neighborhood) yields a $B-V$
color redder by $\sim 0.1$mag and is shown in Fig.~8. 
In the solar neighborhood, older stars are also present, causing 
the horizontal dispersion in color on the fainter edge of the red 
clump in the Hipparcos data.
\bigskip

\noindent
Fig.~4 --- {\bf Top panel:} Calibrated, dereddened $(B-V) - V$ 
color-magnitude diagram for our least crowded field F2.
4666 stars with cosmetically superior point spread functions in 
$B$ and $V$ are shown.  
Isochrones with LMC metallicity ($Z = 0.008$) and 
helium abundance ($Y = 0.25$) with ages 
${\rm log}_{10} {\rm (Age)} = \, $8.4, 8.6, 8.8, 9.0, and 9.4 
from Bertelli \etal\ (1994) are shown 
superposed for two different assumptions for the distance modulus, 
$\mu_{LMC}$ to the LMC. 
{\bf Bottom panel:} Contour representation of the CMD density 
for field F2 in region of the red clump shown superposed on the 
mean locus of the core Helium burning phase as a function of age
as in Fig.~3.  Separate scales for the abscissa of the top and 
bottom panels are indicated.  The region of CMD shown has been divided 
into 25 bins along both the magnitude and color axes.  For this binning, 
contour levels displayed are 12, 10, 8, 6, 4 and 2.5 stars.
A shorter distance modulus of $\mu_{LMC} = 18.3$ provides improved 
fits to the main sequence and red clump. 
\bigskip

\noindent
Fig.~5 --- Calibrated, dereddened $(V-R) - V$ 
color-magnitude diagrams for each field as in Fig~2 
with the isochrones from Fig.~4 overplotted. 
Here, and in what follows, $\mu_{LMC} = 18.3$ 
is assumed.

\bigskip

\noindent 
Fig.~6 --- Same as the bottom panel of Fig~4, but for all of our LMC fields. 
The region of CMD shown has been divided 
into 25 bins along both the magnitude and color axes.  For this binning, 
contour levels displayed are 25, 20, 15, 12, 10, 8, 6, 4 and 2.5 stars.  
The highest contour levels are not always present in the outer fields.

\bigskip

\noindent 
Fig.~7 --- Histograms of main sequence stars for each of our fields 
fields are show on a logarithmic scale.  The number of stars in each 
magnitude bin expected from a Salpeter IMF, taking into account the 
main sequence lifetime and assumed a constant star formation rate 
over the last 1 Gyr, is shown as the solid sloping line.  The 
normalization has been adjusted to achieve a good fit to the 
bulk of the main sequence stars between $16.5 < V < 18.5$.  The 
faintest VRC and main sequence stars used for the 
counting arguments of \S2.3.2 are indicated by the vertical lines 
at $V = $18.5 and 19.5, respectively.  The main sequence counts 
clearly suffer from some incompleteness, whereas the VRC counts do not.
Comparison with expectations from the Salpeter IMF produce the 
main sequence completeness estimates {\it C$_{MS}$\/} indicated 
for each field.

\bigskip

\noindent
Fig.~8 --- {\bf Top panel:} Hipparcos $(B-V) - V$ 
color-magnitude diagram containing over 16000 stars with accurate 
distances and photometry.  A vertical extension can be seen 
clearly on the blueward
side of the red clump with nearly the same color as the peak clump 
density and extending about a magnitude brighter. 
The redder supra-clump is also visible.
Isochrones with solar metallicity (Z = 0.02) 
and ${\rm log}_{10} {\rm (Age)}$ 8.6, 9.0, 9.4, 9.8 and 10.0 are shown 
overplotted.  Note that positions of the blue loops 
again coincide with the vertical extension.  
{\bf Bottom panel:} Contour representation of the CMD density 
in region of the red clump for the solar neighborhood as 
measured by Hipparcos is shown superposed on the mean locus of the 
core He-burning phase as a function of age for stars with metallicity 
and helium abundance 
appropriate to the solar neighborhood ($Z = 0.02$ and $Y= 0.28$). 
Separate scales for the abscissa of the top and 
bottom panels are indicated.  The region of CMD shown has been divided 
into 20 bins along both the magnitude and color axes.  For this binning, 
contour levels displayed are 35, 25, 15, 10, 5 and 2.5 stars.  

\bigskip

\noindent
Fig.~9 --- The same isochrones of LMC metallicity from Fig.~3 are
plotted in color-magnitude diagrams for the region of the red clump. 
The color $C \equiv 0.565*(B- I) + 0.825*(U - V + 1.15)$ is that used 
by Zaritsky \& Lin  (1997) as are the bands on the vertical axis: 
U (top), B (middle), and I (bottom) respectively. 
Reddening vectors assuming an average $E(B-V) = 0.12$ are shown in each panel. 
The blue loops correspond with the feature seen by ZL at a 
dereddened color of $C \approx 2.9$.

\newpage

\begin{table}
\caption[]{Field center, seeing, and reddening determined from foreground 
and main-sequence fitting for each field.}
\begin{flushleft}
\begin{tabular}{lllrrr}
Field & RA (J2000) & Dec (J2000) & Seeing & E(B-V)$_{\rm\textstyle f}$ & E(B-V)$_{\rm\textstyle ms}$\\
\hline
\hline
F1 & 5h 26\arcmin\ 34\arcsec\ & $-$70\deg\ 57\arcmin\ 45\arcsec\ & 1.0\arcsec\ & 0.18 & 0.18\\
F2 & 5h 06\arcmin\ 05\arcsec\ & $-$65\deg\ 58\arcmin\ 03\arcsec\  & 1.3\arcsec\ & 0.08 & 0.12\\
F3 & 5h 14\arcmin\ 44\arcsec\ & $-$68\deg\ 48\arcmin\ 00\arcsec\ & 1.0\arcsec\ & 0.12 & 0.16\\ 
F4 & 5h 23\arcmin\ 00\arcsec\ & $-$66\deg\ 58\arcmin\ 00\arcsec\ & 1.3\arcsec\ & 0.12 & 0.12\\
\hline
\end{tabular}
\end{flushleft}
\end{table}

\newpage

\begin{table}
\caption[]{Position of the clump, relative limits of its vertical extension 
(VRC), and relative position of the supra-clump.}
\begin{flushleft}
\begin{tabular}{lrrrr}
  & F1 & F2 & F3 & F4 \\ 
\hline
\hline
Red Clump & & & & \\
\hline
$V_0$     & 18.67 $\pm$ 0.03 & 18.65 $\pm$ 0.02 &  18.68 $\pm$ 0.03 & 18.70 $\pm$ 0.02   \\
$(B-V)_0$ & 0.87 $\pm$ 0.01  & 0.86 $\pm$ 0.01  &  0.85 $\pm$ 0.01  &  0.88 $\pm$ 0.02\\
$(V-R)_0$ & 0.48 $\pm$ 0.01  & 0.48 $\pm$ 0.01  &  0.48 $\pm$ 0.01  & 0.48 $\pm$ 0.01\\
\hline
VRC & & & & \\
\hline
 $\delta V_0$     & -0.80 $\pm$ 0.05 & -0.82 $\pm$ 0.05 & -1.02 $\pm$ 0.05 & -0.63 $\pm$ 0.10 \\
 $\delta (B-V)_0$ & -0.01 $\pm$ 0.01 &  0.01 $\pm$ 0.01 & 0.01 $\pm$ 0.01 & 0.00 $\pm$ 0.02 \\
 $\delta (V-R)_0$ &  0.00 $\pm$ 0.01 & -0.02 $\pm$ 0.01 & 0.00 $\pm$ 0.01 & 0.00 $\pm$ 0.02\\
\hline
Supra Clump & & & & \\
\hline
 $\delta V_0$     &  -0.82 $\pm$ 0.05 & -0.85 $\pm$ 0.05 & -0.84 $\pm$ 0.05 & -0.86 $\pm$ 0.05\\
 $\delta (B-V)_0$ &  0.24 $\pm$ 0.02 & 0.24 $\pm$ 0.02 & 0.20 $\pm$ 0.02 &0.25 $\pm$ 0.02 \\
 $\delta (V-R)_0$ &  0.10 $\pm$ 0.02 & 0.11 $\pm$ 0.02 & 0.10 $\pm$ 0.02  &0.10 $\pm$ 0.01 \\
\hline
\end{tabular}
\end{flushleft}
\end{table}

\begin{table}
\caption[]{Number counts and relative fractions of the vertical 
extension to the red clump (VRC), supra-clump (SC), 
and red clump (RC) in the individual and composite fields.  
See \S2.1.2 for the operative 
definition used here for each of these regions in the CMD.}
\begin{flushleft}
\begin{tabular}{lrrrrr}
Field & VRC & SC & RC & VRC/RC & SC/RC \\ 
\hline
\hline
F1 &  45 & 92 & 777 &  6\% & 12\% \\
\hline
F2 &  39 & 34 & 261 & 15\% & 13\% \\
\hline
F3 & 97 & 176 &  1074 &  9\% & 16\% \\
\hline
F4 & 19 &  49 & 330 &  6\% & 15\% \\
\hline
Comp & 200 & 351 & 2442 & 8\% & 14\% \\ 
\hline
\end{tabular}
\end{flushleft}
\end{table}

\end{document}